\documentclass[%
 reprint,
 amsmath,amssymb,
 aps,
]{revtex4-2}

\usepackage{graphicx}% Include figure files
\usepackage{dcolumn}% Align table columns on decimal point
\usepackage{bm}% bold math
\usepackage{upgreek}% Non-italic Greek letters
\usepackage{xcolor}% Used to set colors to, e.g., text (comments to myself or others). 
\begin{document}

\title{Designing gate operations for single ion quantum computing \\in rare-earth-ion-doped crystals}% Force line breaks with \\

\author{Adam Kinos}
\email{adam.kinos@fysik.lth.se}
\author{Lars Rippe}
\author{Stefan Kr\"{o}ll}
\author{Andreas Walther}

\affiliation{%
 Department of Physics, Lund University, P.O. Box 118, SE-22100 Lund, Sweden
}%

\date{\today}% It is always \today, today,

\begin{abstract}
Quantum computers based on rare-earth-ion-doped crystals show promising properties in terms of scalability and connectivity if single ions can be used as qubits. Through simulations, we investigate gate operations on such qubits and discuss how gate and system parameters affect gate errors, the required frequency bandwidth per qubit, and the risk of instantaneous spectral diffusion (ISD) occurring. Furthermore, we examine how uncertainties in the system parameters affect the gate errors, and how precisely the system needs to be known. We find gate errors for arbitrary single-qubit gates of $2.1\cdot10^{-4}$ when ISD is not considered and $3.4\cdot10^{-4}$ when we take heed to minimize it. Additionally, we construct two-qubit gates with errors ranging from $5\cdot 10^{-4} \rightarrow 3\cdot 10^{-3}$ over a broad range of dipole-dipole interaction strengths.
\end{abstract}

\maketitle

% ---------------------------------------------------------------------------
\section{\label{sec:intro}Introduction}
Quantum computing is imagined on many different platforms, such as superconducting qubits \cite{Clarke2008, Arute2019, Kjaergaard2020}, and trapped ions \cite{Bruzewicz2019, Lekitsch2017, Kaushal2020}, but also using single ions in the solid state system of rare-earth-ion-doped crystals. The rare-earth platform is versatile and has shown promise in fields related to quantum computing, such as quantum memories \cite{Nilsson2005a, Kraus2006, Hetet2008, Riedmatten2008, Afzelius2010, Hedges2010, Beavan2012a, Sabooni2013a, Dajczgewand2014, Guendogan2015, Jobez2015, Schraft2016}, conversion between optical and microwave signals \cite{OBrien2014, Williamson2014, FernandezGonzalvo2019}, and single photon sources \cite{Dibos2018}. Much work has also been devoted to quantum computing \cite{Pryde2000, Ichimura2001, Ohlsson2002, Nilsson2002, Wesenberg2003, Wesenberg2007, Walther2009a, Walther2015, Ahlefeldt2020, Grimm2021, Hizhnyakov2021}, which is the topic that here is investigated through simulations. Since the topic is rather complex, we start by providing a general overview of the system. 

The qubit is encoded in the ground hyperfine states, which can have lifetimes of days \cite{Konz2003} and coherence times of hours \cite{Zhong2015}, while gate operations are performed using optical transitions to an excited state with coherence times of up to milliseconds \cite{Equall1994, Sun2002}. Individual qubits are addressed in frequency space, where each qubit occupies a specific frequency channel, of which there can be many thanks to the large inhomogeneity in the optical transitions \cite{Konz2003}. The spectral addressing makes it possible for the ions to be packed at nanometer spacing in three dimensions, thus providing very high qubit densities. This in turn allows for strong dipole-dipole interactions even between qubits that are not nearest neighbours \cite{Ohlsson2002, Nilsson2002, Longdell2004a, Longdell2004, Ahlefeldt2013b}, leading to high connectivity. The dipole-dipole interaction might also be used when reading out the state of qubits via co-doped readout ions \cite{Wesenberg2007}. The optical transitions simplifies the process of transferring quantum information into flying qubits, which opens up the possibility to perform remote entanglement and construct quantum networks. 

However, due to the difficulty of detecting single ions, most results so far have been limited to use the ensemble approach where the qubit consists of many rare-earth-ions, and where the resulting inhomogeneities complicate the gates and limit the fidelities \cite{Longdell2004a, Longdell2004, Rippe2005, Rippe2008, Roos2004, Tordrup2007, Yan2019, Yan2021}. Furthermore, this ensemble approach has been shown to scale poorly \cite{Wesenberg2007}, but might still be useful, e.g., when using stoichiometric materials where small quantum processors have been envisioned \cite{Ahlefeldt2020}. Fortunately, recent progress in single-ion detection \cite{Utikal2014, Xia2015, Kolesov2018, Zhong2018, Raha2020, Kindem2020} paves the way for single-ion rare-earth quantum computing. 
In order to focus the research community, a roadmap for rare-earth quantum computing using single-ion qubits has now been developed \cite{Kinos2021}. One important step in this process is choosing a good system and designing high fidelity gate operations in a scalable and realistic way. This is complicated since each of the system and gate parameters affect the gate operations and the properties of the quantum computer in multiple different ways. For example, the gate duration affects the gate error in three different ways, but it also sets an upper limit to the speed of the quantum computer, and affects the required frequency bandwidth per qubit thus potentially limiting the number of qubits available in the quantum computer. 

In the present work we design qubit operations that take the known complexities into account and explain what effects different parameter choices have. Our intention is that the detailed analysis and discussion of single- and two-qubit operations provided in this work can form a foundation for studying interactions and operation fidelities involving larger number of qubits. In this way it can further develop the possibilities to reliably analyze scalability of rare-earth-ion-based quantum computing.

\begin{figure*}
\includegraphics[width=\textwidth]{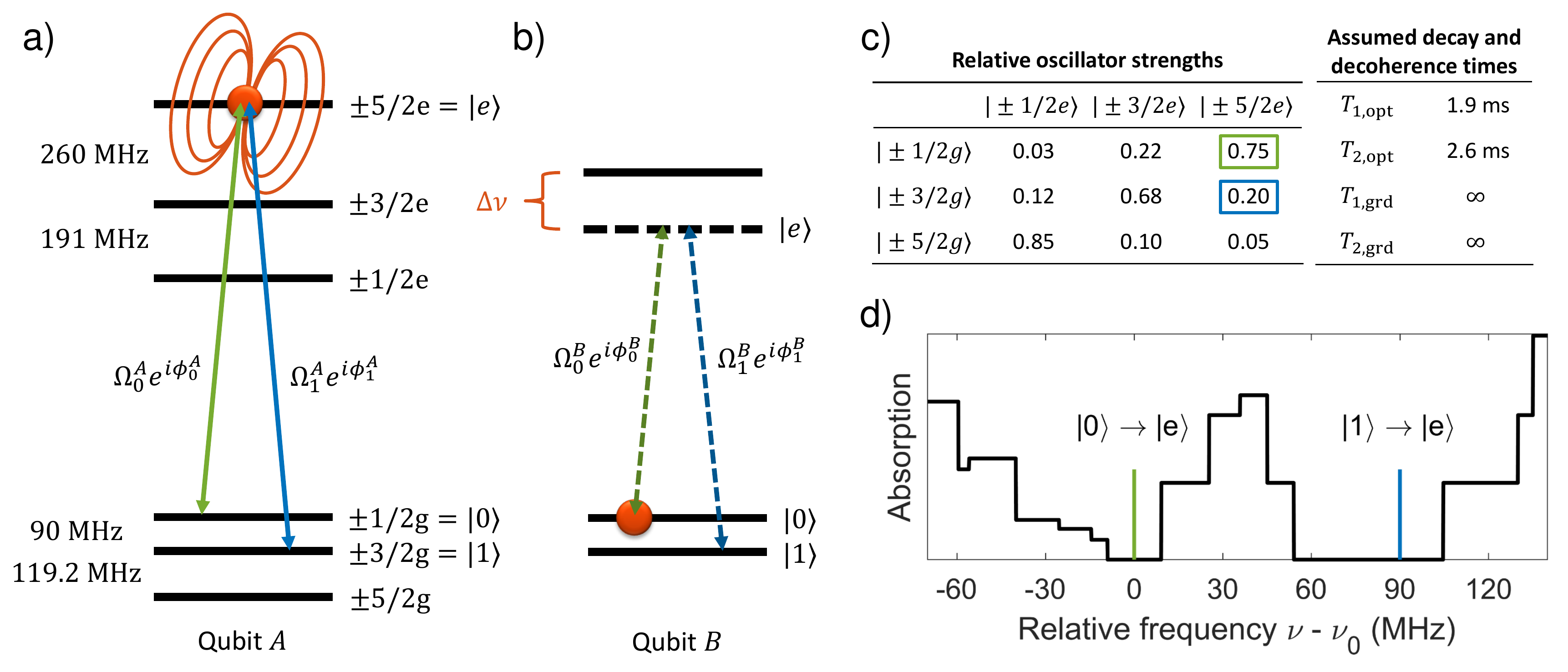}
\caption{\label{fig:Enery_levels}a) Energy level structure for $^{153}$Eu:Y$_2$SiO$_5$ site 1 \cite{Sun2005}, including the two optical driving fields with strengths $\Omega_0^A$ and $\Omega_1^A$, and phases $\phi_0^A$ and $\phi_1^A$, respectively. All simulations assume a zero magnetic field in the sense that the hyperfine levels are doubly degenerate, i.e., $|$+1/2g$\rangle$ overlaps with $|$-1/2g$\rangle$ and will therefore be treated as one single level, $|\pm$1/2g$\rangle$. b) Relevant energy levels for a second qubit, qubit $B$. Its optical transitions are detuned from qubit $A$ due to an inhomogeneous broadening which is concentration dependent and can be up to hundred GHz broad \cite{Konz2003}. The two qubits interact through a dipole-dipole interaction which shifts the optical transitions of one qubit when the other is excited. The shift, $\Delta \nu$, is proportional to $1/r^3$, where $r$ is the distance between the qubits. c) The relative oscillator strengths of the different transitions \cite{Lauritzen2012}, and the assumed life-, $T_1$, and coherence, $T_2$, times for the optical and ground transitions \cite{Equall1994}. The ground state times are assumed infinite since they can be much longer compared to their optical counterparts \cite{Konz2003, Alexander2007, Arcangeli2013, Zhong2015}. d) The optimal transmission windows in the inhomogeneously broadened absorption profile surrounding the frequencies of the two qubit transitions $|0\rangle \rightarrow |e\rangle$ and $|1\rangle \rightarrow |e\rangle$, indicated by the green and blue lines, respectively,, where the former transition has a frequency of $\nu_0$.}
\end{figure*}

An overview of the sections and content of the paper is now provided. In Sec. \ref{sec:sq_gates} the system and the single-qubit gate operations used are described and motivated. The connection between different error sources and system and gate parameters are discussed, and gate parameters giving low gate errors are provided. The error sources we investigate are categorized in three different ways: \textit{decay and decoherence}, which include fundamental and unknown error sources; \textit{internal crosstalk}, which we define as error sources that only depend on the properties and control of an individual qubit, and in this work we assume that internal crosstalk only occurs when gate operation pulses off-resonantly drive other transitions in the qubit than the intended ones; and finally, \textit{external crosstalk}, which we define as error sources that depend on properties external to an individual qubit, e.g., errors occurring when pulses intended to drive one qubit also off-resonantly drive the transitions in another qubit, or instantaneous spectral diffusion (ISD) where a non-qubit ion is unintentionally excited and cause an instantaneous frequency shift of the optical transitions of the qubit. Sec. \ref{sec:pulses} investigates how gate operations on one qubit affects another qubit, how this scales when adding more qubits, and how it connects to the frequency bandwidth per qubit and hence the number of potential qubits in the quantum computer. Following this, two types of two-qubit gate operations, the blockade and interaction gates, are analyzed in Sec. \ref{sec:tq_gates}. Then, Sec \ref{sec:uncert} determines how precisely the system parameters need to be known with respect to uncertainties and fluctuations. In this section we also briefly discuss another external crosstalk effect, namely spectral diffusion \cite{Klauder1962, Boettger2006, Thiel2014a}, which is different from ISD since the latter only occurs when other non-qubit ions are excited by the gate operation pulses. Finally, the paper is concluded in Sec. \ref{sec:conc}.

% ---------------------------------------------------------------------------
\section{\label{sec:sq_gates}Single-qubit gate operations}
In order to investigate how high the fidelity of single- and two-qubit gate operations can be in a realistic case using single ions, we start by deciding on a system. Here we investigate site 1 of $^{153}$Eu:Y$_2$SiO$_5$, whose relevant properties such as energy level structure, relative oscillator strengths, and life- and coherence times used in the calculations can be seen in Fig. \ref{fig:Enery_levels}. We pick $|\pm$1/2g$\rangle = |0\rangle$ and $|\pm$3/2g$\rangle = |1\rangle$ as our qubit system. Motivations behind these choices are provided later in this section.

Single-qubit (SQ) gate operations can be performed in several different ways, e.g., see references \cite{Bauer1984, Wesenberg2003, Roos2004, Theis2018, Setiawan2021}. In this work we focus on arbitrary SQ gate operations performed using 2 two-color optical pulses resonant with the transitions from the qubit levels to the excited state $|\pm$5/2e$\rangle = |e\rangle$, roughly following the scheme presented in \cite{Roos2004}. Since we investigate a resonant single-ion qubit, the dark state compensation pulses used in \cite{Roos2004}, which are designed to fix a phase error for ions that are optically detuned from the pulses, are no longer necessary and are not used. Furthermore, the hyperbolic secant pulse shape in \cite{Roos2004}, which achieves robustness against certain fluctuations at the cost of a longer gate duration, is changed to a cut Gaussian envelop to achieve the required pulse area in a shorter duration. For completeness, we describe the SQ gate operation procedure here. 

\begin{figure*}
\includegraphics[width=\textwidth]{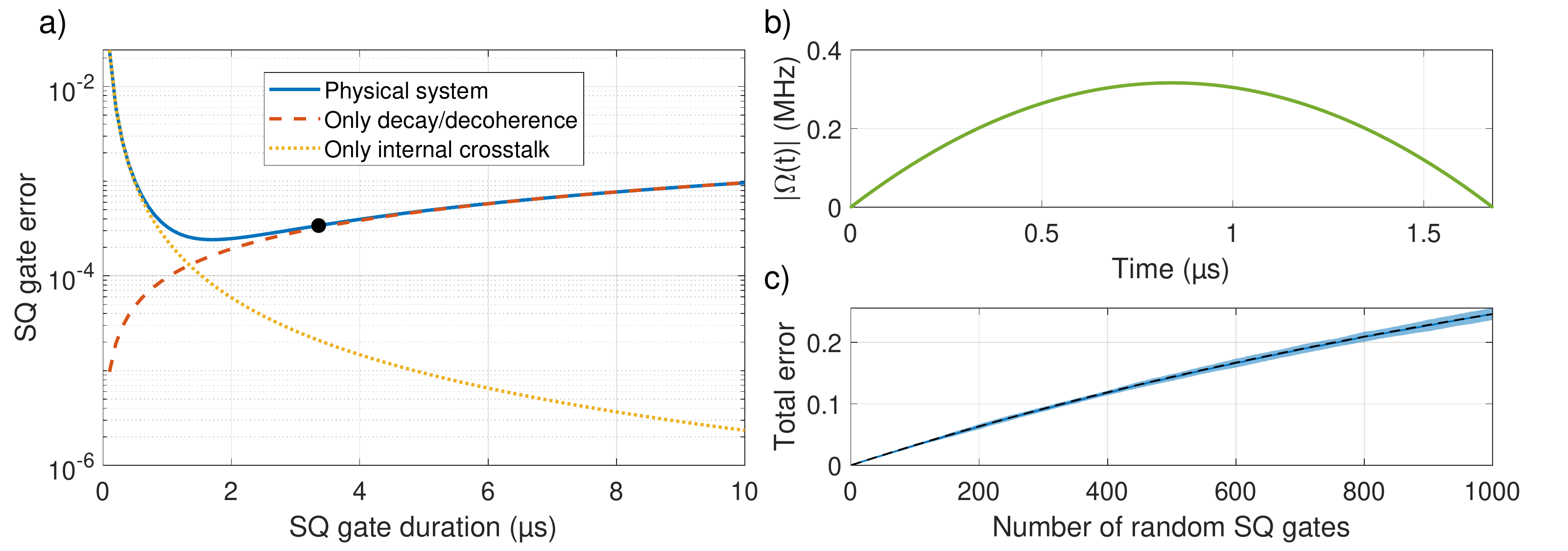}
\caption{\label{fig:SQ_eval}a) Single-qubit (SQ) gate error as a function of the SQ gate duration for a physical system (solid blue), a system with only decay and decoherence (dashed red), and a system with only internal crosstalk (dotted yellow). As can be seen, short (long) gate durations lead to errors mainly due to internal crosstalk (decay and decoherence). The pulse shape of the gate operation is defined in Eq. \ref{eq:Gaussian} with $t_g / \sigma = 0.4$, where $2t_g$ is the SQ gate duration. The black dot shows the total gate duration for the optimized parameters used to simulate arbitrary SQ gate operations in this work, which means that our gate is mostly limited by decay and decoherence. b) Shows the Rabi frequency envelope of the optimized pulse with parameters: $t_g = 1.68$ $\upmu$s and $\sigma = 4.16$ $\upmu$s in Eq. \ref{eq:Gaussian}, with a total gate duration of $2t_g = 3.36$ $\upmu$s. c) Shows the total SQ error as a function of the number of randomized arbitrary SQ gate operations using the optimized gate parameters listed above and performed on the system depicted in Fig.~\ref{fig:Enery_levels}. The simulation is repeated 100 times, and the solid blue line shows the average error whereas the lighter blue region shows the standard deviation. The dashed black line is a fit based on Eq. \ref{eq:SQ_err} with an average error rate of $p \approx 3.4\cdot10^{-4}$.}
\end{figure*}

First, a two-color pulse with Rabi frequencies $\Omega_0(t) e^{i\phi_0}$ and $\Omega_1(t) e^{i\phi_1}$, that have a relative phase of $\phi = \phi_1 - \phi_0$, is sent in. Both components have the same cut Gaussian pulse shape, $\Omega_0(t) = \Omega_1(t) = \Omega(t)$: 
\begin{equation}\label{eq:Gaussian}
  \Omega(t) =
    \begin{cases}
      C_1\cdot \text{exp}(-\frac{(t-t_g/2)^2}{2\sigma^2}) - C_2 & 0 \leq t \leq t_g\\
      0 & \text{otherwise}
    \end{cases}       
\end{equation}

where $t_g$ is the cut-off pulse duration, $\sigma$ is the standard deviation of the Gaussian, $C_1$ is chosen so that a pulse area of $\pi/\sqrt{2}$ is achieved, and $C_2$ enforces the shape to start and end at zero \cite{Bauer1984, Motzoi2009}. The qubit system now has two superpositions, called bright and dark, which are coupled or uncoupled, respectively, to the excited state. These superpositions are defined as follows \cite{Roos2004}: 
\begin{eqnarray}
    |B\rangle = \frac{1}{\sqrt{2}} \left( |0\rangle + e^{-i\phi} |1\rangle \right) \\
    |D\rangle = \frac{1}{\sqrt{2}} \left( |0\rangle - e^{-i\phi} |1\rangle \right)
\end{eqnarray}

The interaction strength with bright state is increased by a factor $\sqrt{2}$, and since the pulse area of the cut Gaussian is chosen to be $\pi/\sqrt{2}$, the first pulse performs a $\pi$ rotation, thus exciting $|B\rangle$ to $|e\rangle$. 

The second two-color pulse is identical to the first, except for a phase change of the driving fields, 
\begin{eqnarray}\label{eq:theta}
    \phi_0^{new} = \phi_0 + \pi - \theta \nonumber\\
    \phi_1^{new} = \phi_1 + \pi - \theta
\end{eqnarray}

This deexcites $|e\rangle$ back to $|B\rangle$, but with an added phase of $e^{i\theta}$ compared to the dark state, $|D\rangle$. By picking various choices of $\phi$ and $\theta$ any SQ rotation can be performed, keeping in mind that one can always redefine the phases of subsequent pulses to perform a virtual $z$ rotation \cite{Mckay2017}. 

The simulations in this article numerically solve the Lindblad master equation \cite{Manzano2020} with either $6$ or $36$ energy levels depending on if the simulation is for one or two qubits, see Appendix \ref{app:simulations} for more information. Unless otherwise specified, the gate errors provided are the average error when averaging over six initial states for each qubit: $|0\rangle$, $|1\rangle$, $|0\rangle + |1\rangle$, $|0\rangle - |1\rangle$, $|0\rangle + i|1\rangle$, $|0\rangle - i|1\rangle$ \cite{Bowdrey2002}, where we disregard any normalization factors, and six gate operations: $I$, $X$, $\sqrt{X}$, $\sqrt{-X}$, $\sqrt{Y}$, $\sqrt{-Y}$, where $I$ is the identity operator, and $X$ and $Y$ are the Pauli operators \cite{QuantumLogicGateWiki}. The error, $\epsilon$, is calculated using:
\begin{equation}
    \epsilon = 1 - \langle \Psi | \rho | \Psi \rangle
\end{equation}

where $\rho$ is the achieved density matrix from the simulation and $\Psi$ is the desired target state. 

We now investigate three error sources that affect the SQ gate error: decay and decoherence; internal crosstalk; and instantaneous spectral diffusion (ISD). In Fig. \ref{fig:SQ_eval}a one can see how the average SQ gate error depend on the gate duration as well as the main source of the error. Long gate durations lead to errors due to decay and decoherence. Short gate durations lead to internal crosstalk as the pulses off-resonantly interact with other energy levels in the qubit than the intended ones. This can be in the form of state leakage where population is transferred to another energy level, which scales as $1/\delta^2$ where $\delta$ is the detuning to a nearby unwanted transition, or phase errors due to an AC Stark shift, which scales as $1/\delta$ \cite{Chen2016, Wood2018}. 

Instantaneous spectral diffusion (ISD) is another error source in rare-earth-ion-doped crystals where a non-qubit ion interacts with a qubit ion in an unpredicted way via the dipole-dipole interaction explained in Fig. \ref{fig:Enery_levels}b. Here the pulses intended to perform a gate operation on the qubit ion also excites a non-qubit ion (potentially off-resonantly) which through the dipole-dipole interaction shifts the resonant frequencies of the qubit ion. This can lead to either phase errors on the qubit or an unwanted entanglement between the qubit and the non-qubit ion. The effect on the SQ gate error from this interaction depends on both how detuned the non-qubit ion is from the pulses and the distance between the non-qubit and qubit ion, since the latter determines the dipole-dipole interaction strength. ISD is therefore a stochastic error which depend on the properties of the non-qubit ions surrounding a particular qubit ion. 

Using this knowledge, one can now motivate our choice of system. $^{153}$Eu:Y$_2$SiO$_5$ site 1 have long life- and coherence times making it possible to achieve low gate errors even for relatively long gate durations. Furthermore, the level splittings are relatively large, thus minimizing internal crosstalk. Lastly, the qubit and excited states $|0\rangle$, $|1\rangle$, and $|e\rangle$, are picked because the relative oscillator strengths for those transitions are relatively high, $0.75$ and $0.20$, respectively, and the frequency separation to other transitions are relatively large, and both these properties also help in minimizing internal crosstalk. 

Furthermore, one way to minimize the risk of ISD occurring is to isolate the qubit ion in frequency space from non-qubit ions which otherwise may be excited and affect the gate operation. This can be done using spectral hole burning techniques \cite{Nilsson2002, Nilsson2004, Rippe2005, Lauritzen2012}, where semipermanent transmission windows are burned into the inhomogeneously broadened rare-earth-ion ensemble, see Fig. \ref{fig:Enery_levels}d. One can calculate the largest possible widths of such windows by iterating through all inhomogeneously broadened ions and placing them in the ground state whose transitions to all excited states are as far away as possible from the frequencies of the two qubit transitions. These widths depend on which qubit system is picked, and wide windows reduce the risk of ISD more than narrow windows since it forces the non-qubit ions to be further detuned from the gate operation pulses. For our choice of system, the empty regions surrounding the two optical transitions, $|0\rangle \rightarrow |e\rangle$ and $|1\rangle \rightarrow |e\rangle$, range from $-9.0$ MHz to $9.1$ MHz, and $-35.9$ MHz to $14.6$ MHz, measured from the center of the respective transitions. The fact that the second window is wider is good since that transition has a lower relative oscillator strength, and therefore requires stronger fields to drive with the same Rabi frequency. Outside of these empty regions there are a lot of non-qubit ions that if excited can cause ISD which may increase the gate error. Note that in a real experimental setup the transmission windows might have some residual absorption. Furthermore, depending on how the qubit ion is initialized there may be multiple other ions that are also transferred back into the transmission windows. Such ions can possible be removed again, e.g., using the technique described in reference \cite{Wesenberg2007}. Neither of these effects are considered further in this article, and the focus instead lies on minimizing the impact on the ions outside the transmission windows.

Note that ISD can potentially be mitigated by lowering the concentration of the dopant in randomly doped crystals. However, one should keep in mind that doing so also increases the average distance between qubit ions, thus making it more difficult to find qubits that interact strongly. ISD might also be completely avoided if, e.g., ion implantation is used to directly implant the qubit ions into a crystal, thus circumventing the problem of all the non-qubit ions that exist in randomly doped systems \cite{GrootBerning2019}. However, presently implanted techniques have other issues, e.g., the coherence properties of the implanted ions are sometimes worse than in bulk materials \cite{Kinos2021}, and so more work is required on such methods.

A detailed investigation of how ISD quantitatively affects the SQ gate operation in a stochastic manner is beyond the scope of this work. Instead we design two different sets of gate pulses; one set which can be used if no ISD is present, and another set which takes heed for ISD by minimizing the effect on all the non-qubit ions just outside the empty transmission windows. 

When no ISD is present one can optimize for the lowest SQ gate error achievable in the physical system curve of Fig. \ref{fig:SQ_eval}a, which yields an average SQ gate error of roughly $2.4\cdot10^{-4}$. However, there are more complicated gate schemes that can be beneficial to use, e.g., DRAG \cite{Motzoi2009, Gambetta2011, Motzoi2013, Theis2018}, which reduces the error due to internal crosstalk by modifying the driving pulses, or accelerated adiabatic pulses \cite{Ribeiro2019, Setiawan2021}. A slightly lower average SQ gate error of $2.1\cdot10^{-4}$ was found by using a DRAG off-quadrature compensation; $\Omega_\text{off}(t) = \alpha_y\frac{d\Omega(t)}{dt}$, with gate parameters of $\alpha_y = -1.46\cdot10^{-9}$, $t_g = 0.77$ $\upmu$s, and $\sigma = 1.4$ $\upmu$s. Note that while these values gave the lowest SQ gate error in the optimization procedure, similar parameters exist that yield roughly the same error, e.g., $\alpha_y = -2\cdot10^{-9} \rightarrow -1\cdot10^{-9}$, $t_g = 0.65$ $\upmu$s $\rightarrow 0.85$ $\upmu$s, and $\sigma = 0.5$ $\upmu$s $\rightarrow 1.5$ $\upmu$s. For more information about the optimization procedure used to find these gate parameters, see Appendix \ref{app:optimization}. 

The focus of this work, however, is on finding and investigating gate parameters that take heed for ISD as well, so the rest of this work is pursuing that task. Now the optimization of the gate parameters simultaneously tries to achieve the two partially competing goals of finding the lowest SQ gate error while also minimizing the impact on the non-qubit ions at the edges of the transmission windows to reduce the risk of ISD occurring. These goals are partially competing since the former sometimes benefits from shorter gate durations, which lead to less decay and decoherence, whereas the latter benefits from longer gate durations which have lower frequency bandwidths. Again, more information about the optimization procedure can be found in Appendix \ref{app:optimization}. 

The optimization led to pulse parameters of $t_g = 1.68$ $\upmu$s and $\sigma = 4.16$ $\upmu$s, and had an average SQ gate error of $3.4\cdot 10^{-4}$. Thus, our choice to find gate parameters that also take heed for ISD comes at the cost of a slightly higher SQ gate error compared to when such considerations were not taken. DRAG compensation gave no significant improvement this time since the largest error of this gate stem from decay and decoherence, which can be seen in Fig. \ref{fig:SQ_eval}a, where the result for these optimized gate parameters are indicated by the black dot. In general this choice corresponds to choosing a pulse duration where the effect on ions just outside the transmission windows was roughly equal to the SQ gate error of the qubit. Note again that the values of the parameters could be tuned in a relatively large range and still provide similar optimization scores, e.g., $t_g = 1.5$ $\upmu$s $\rightarrow 2$ $\upmu$s, and $\sigma > 1$ $\upmu$s. The Rabi frequency envelope for the optimized gate parameters can be seen in Fig. \ref{fig:SQ_eval}b. The optimization led to an almost quadratic pulse shape, thus finding it optimal to deliver the required pulse area with as low maximum Rabi frequency as possible for the given pulse duration while still being forced to maintain the shape of a cut Gaussian. 

Note that the dark state compensation pulses used in \cite{Roos2004} could be useful in this case as they minimize the impact of the gate operation on the detuned non-qubit ions. However, when investigated, the doubling of the gate duration required to also perform the dark state compensation pulses resulted in larger errors than if they were not used. 

In order to benchmark this SQ gate operation we successively perform 1000 random gates by randomizing the phases of the gate pulses $0\leq\phi\leq2\pi$ and $0\leq\theta\leq\pi$, whose definitions are described in the text surrounding Eqs. \ref{eq:Gaussian}-\ref{eq:theta}. We then look at the total error as a function of the number of gate operations. In order to build statistics this is repeated 100 times and the average total error and its standard deviation can be seen in Fig. \ref{fig:SQ_eval}c. 

We also estimate the average error per gate, $p$, by assuming that the gate can either introduce a new error or fix a previous error with the same probability: 
\begin{equation}\label{eq:SQ_err}
    \epsilon_n = \epsilon_{n-1} + p(1-\epsilon_{n-1}) - p\epsilon_{n-1}
\end{equation}

where $\epsilon_n$ is the total error obtained after $n$ gate operations. The average error per gate is $p\approx 3.4\cdot10^{-4}$, and the result of the simple error model can be seen in the dashed black line of Fig. \ref{fig:SQ_eval}c. 

% ---------------------------------------------------------------------------
\section{\label{sec:pulses}Off-resonant driving by pulses intended to drive other qubits}
\begin{figure}
\includegraphics[width=\columnwidth]{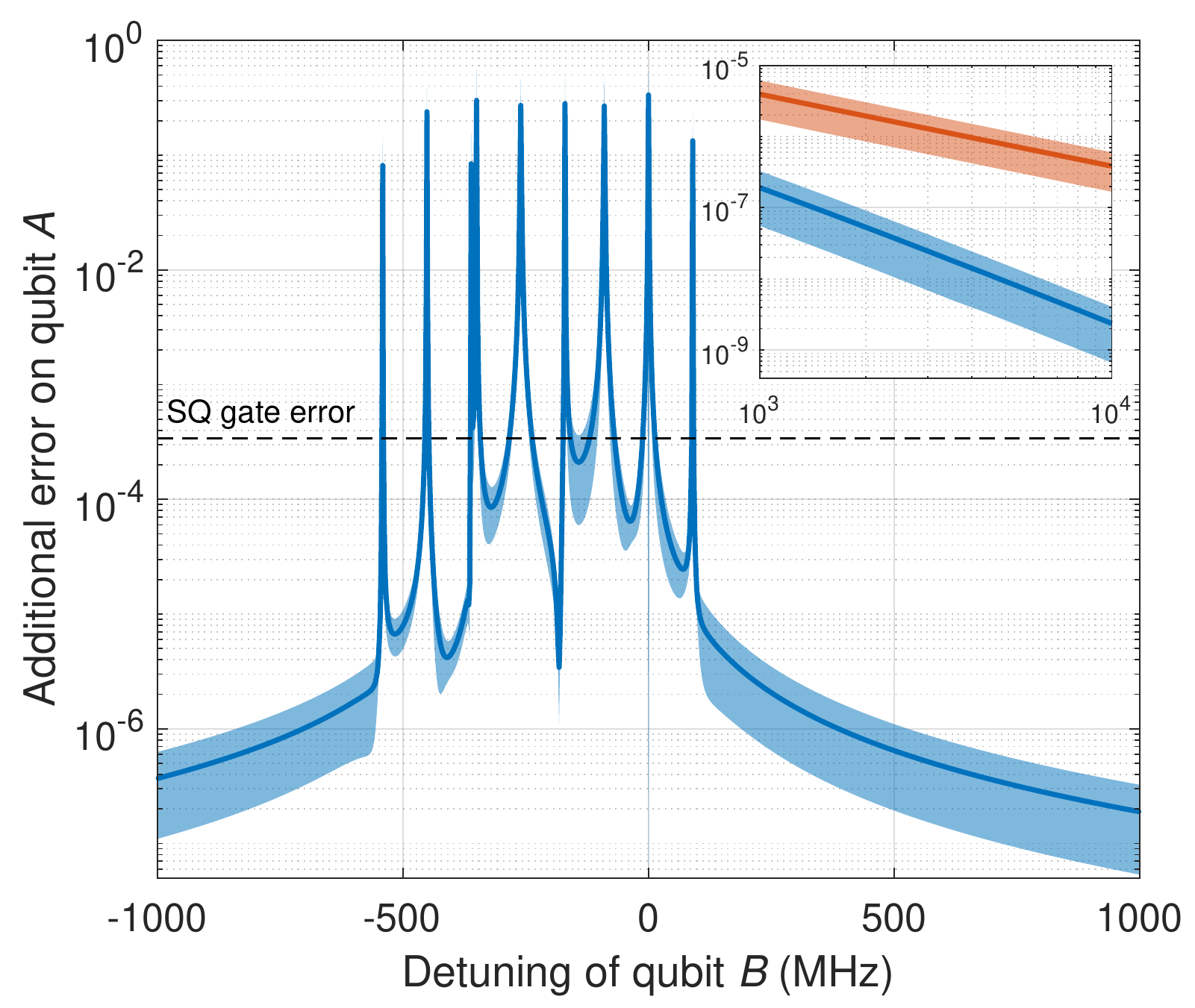}
\caption{\label{fig:Error_A_detuning_B}The pulses intended to perform a SQ gate operation on qubit $B$ also drive transitions in qubit $A$. This introduces additional errors on qubit $A$, which are shown as a function of the detuning, $\Delta$, between qubit $B$'s and qubit $A$'s optical transitions. The error for each detuning is averaged over the previously defined six initial states of qubit $A$ and six gate operations performed on qubit $B$. The solid lines show the average additional error and the lighter regions show the standard deviation of these 36 combinations of initial states and gates. The dashed black line shows the average SQ gate error of $3.4\cdot 10^{-4}$. The inset shows the additional error for large positive detunings, where the gate operation on qubit $B$ is performed simultaneously as no gate (blue data) or a NOT gate (red data) is performed on qubit $A$.}
\end{figure}

The inhomogeneous broadening of the optical transitions allows for control of different qubits simply by changing the laser frequency. However, when scaling up the system to more qubits it is important that a SQ gate operation on one qubit, qubit $B$, doesn't add a significant error to other qubits, e.g., qubit $A$. This can happen since the pulses intended to perform a SQ gate operation on qubit $B$ also off-resonantly drive the transitions in qubit $A$. The additional error due to this interaction is investigated in Fig. \ref{fig:Error_A_detuning_B} as a function of the optical detuning $\Delta$ of qubit $B$ from qubit $A$, where a zero optical detuning means that the two qubits have identical transition frequencies. 

The spikes in the graph occur when the pulses intended to drive either $|0\rangle \rightarrow |e\rangle$ or $|1\rangle \rightarrow |e\rangle$ in qubit $B$ also resonantly drives a transition from $|0\rangle$ or $|1\rangle$ in qubit $A$, thus resulting in large errors. If we also consider the fact that qubit $A$ might be reinitialized during operation and therefore uses the auxiliary state $|\pm 5/2g\rangle$, additional spikes appear when the pulses resonantly drives a transition from the auxiliary state as well, which occur at following optical detunings: -331.8 MHz, -241.8 MHz, -140.8 MHz, -50.8 MHz, 119.2 MHz, and 209.2 MHz. 

For our choice of system and gate parameters we must separate our qubits by more than roughly $600$ MHz to not introduce significant additional errors when performing gate operations on other qubits. Note that although an optical detuning of, e.g., $\Delta = 250$ MHz doesn't introduce a significant additional error on qubit $A$ when qubit $B$ is driven, it does introduce an additional error on qubit $B$ when qubit $A$ is driven, since from the perspective of qubit $B$, qubit $A$ has an optical detuning of $-\Delta = -250$ MHz. Note also that the choice of a system with large level splittings in order to minimize internal crosstalk has a potential drawback here, since its transitions are far apart and therefore might result in a larger bandwidth per qubit, thus limiting the number of qubit channels within the inhomogeneous absorption profile. Similarly, using a system with lower coherence time, or picking weaker qubit transitions or using shorter gate pulses, two choices which require stronger fields, may also result in a larger bandwidth per qubit. 

In the main part of Fig. \ref{fig:Error_A_detuning_B} a SQ gate operation is only performed on qubit $B$ while qubit $A$ remains idle. This corresponds to the case when the SQ gate operations run sequentially on different qubits, which can be done without introducing large errors on the idle qubits thanks to the long coherence times of the ground hyperfine states in rare-earth-ion-doped crystals. Note that running sequentially still leads to longer execution times of algorithms compared to when running gate operations in parallel. However, when running in parallel the additional error on qubit $A$ scales differently with detuning. For large positive optical detunings, $\Delta \geq 1$ GHz, this scaling can be seen in the inset of Fig. \ref{fig:Error_A_detuning_B}. When running sequentially (blue data) the error decreases by roughly $1/\Delta^2$ compared to $1/\Delta$ when running in parallel (red data).

\begin{figure*}
\includegraphics[width=\textwidth]{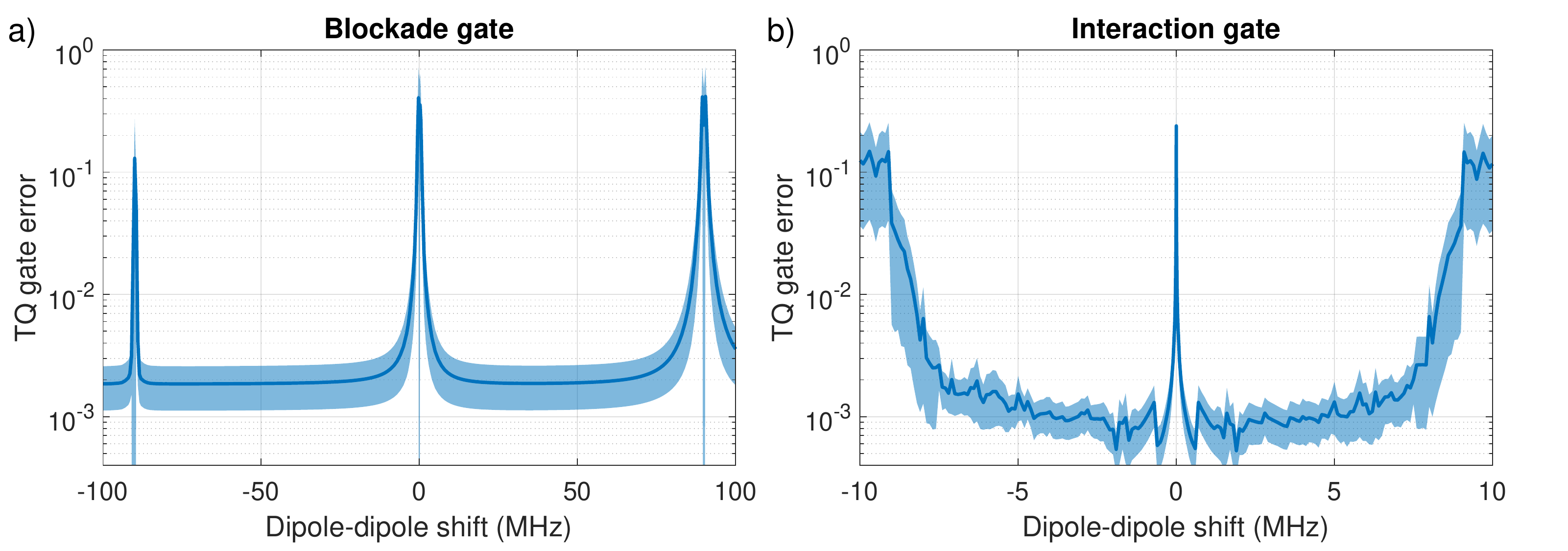}
\caption{\label{fig:TQ_eval}Shows the two-qubit (TQ) gate error as a function of the dipole-dipole interaction shift, $\Delta \nu$, between two qubits for a) a blockade gate utilizing the dipole blockade mechanism to perform control-arbitrary TQ gates, and b) an interaction gate utilizing the exact dipole-dipole shift in order to add an additional phase of $\pi$ to the $|00\rangle$ TQ state. $\Delta \nu$ scales with $1/r^3$, where $r$ is the distance between the two qubits. The solid blue lines show the average error and the lighter blue regions show the standard deviation when averaging over six initial states for each qubit, and for the blockade gate it also averages over six different gate operations performed on the target qubit. }
\end{figure*}

It is also important to investigate how the additional error scales when qubit $A$ is driven off-resonantly by many pulses intended to drive multiple different qubits. When running sequentially the additional error from many qubits each separated by the order of $1$ GHz saturates relatively quickly thanks to the $1/\Delta^2$ dependence. However, when running in parallel the effect is more difficult to analyze. For example, for large negative optical detunings the pulses intended to drive qubit $B$, which also off-resonantly drive qubit $A$, actually compensates for a small part of the gate error otherwise obtained when performing the NOT operation on qubit $A$. Thus resulting in a slightly lower SQ gate error on qubit $A$ when qubit $B$ is driven simultaneously. However, from the perspective of qubit $B$, qubit $A$ has a large positive detuning, which from the inset of Fig. \ref{fig:Error_A_detuning_B} we already know leads to an additional error, this time for the operation on qubit $B$. Fortunately, even when assuming the worst case where the additional error only decreases as $1/\Delta$, it only grows as the logarithm of the number of additional qubits (each separated by the order of $1$ GHz). The effect on the SQ gate error can therefore still be kept under relatively good control, and gate operations can be performed in parallel with only a limited impact on gate errors if the decreased execution time of the algorithm is important.

However, in addition to the fact that the pulses intended to drive qubit $B$ also off-resonantly drives qubit $A$, if one run SQ gate operations in parallel they can also interact via the dipole-dipole interaction explained in Fig. \ref{fig:Enery_levels}a-b. When the qubits are sufficiently far apart in space, this interaction is weak and introduces no significant additional error and the gate operations can run in parallel. But, when the qubits are too close in space the SQ gate operations must be performed sequentially in order to maintain a low SQ gate error.

% ---------------------------------------------------------------------------

\section{\label{sec:tq_gates}Two-qubit gate operations}
We now turn our investigation toward two different methods of performing two-qubit (TQ) gate operations, namely, a blockade gate which through the dipole blockade mechanism can perform control-arbitrary gates \cite{Ohlsson2002, Wesenberg2007}, and an interaction gate which uses the exact dipole-dipole shift in order to evolve a phase on a specific TQ state \cite{Longdell2004, Rao2014}. In these simulations we assume that the qubits are placed sufficiently far apart in frequency so that the pulses driving one qubit doesn't affect the other. 

The blockade gate is constructed in the following way, using qubit $A$ as control and qubit $B$ as target:

\begin{enumerate}
  \item Excite qubit $A$ on $|0\rangle \rightarrow |e\rangle$, thus shifting the resonant frequencies of qubit $B$ through the dipole-dipole interaction explained in Fig. \ref{fig:Enery_levels}a-b, if qubit $A$ was in $|0\rangle$ 
  \item Apply an arbitrary SQ gate pulse sequence on qubit $B$
  \item Deexcite qubit $A$ via the transition $|0\rangle \rightarrow |e\rangle$ with an additional phase of $\pi$, thus deexciting it on the same path as it was excited
\end{enumerate}

The operations on qubit $A$ uses single-color pulses that only interacts with the $|0\rangle \rightarrow |e\rangle$ transition, and new gate parameters of the cut Gaussian defined in Eq. \ref{eq:Gaussian} were optimized for this case: $t'_g = 2.17$ $\upmu$s, $\sigma' = 6.75$ $\upmu$s, and $C_1'$ is chosen so that a pulse area of $\pi$ is achieved. The apostrophe is used to differentiate these parameters from those used for a SQ gate operation. This gives a total TQ gate duration of $t'_g+2t_g+t'_g = 7.7$ $\upmu$s. 

The gate operation works when the dipole-dipole shift, $\Delta \nu$, is large enough to render the arbitrary gate operation on qubit $B$ ineffectual when qubit $A$ is in $|0\rangle$. The average TQ gate error when averaging over the previously defined six initial states for each qubit and six gate operations on the target, can be seen in Fig. \ref{fig:TQ_eval}a. 

The difference frequency between the transitions $|0\rangle \rightarrow |e\rangle$ and $|1\rangle \rightarrow |e\rangle$ is 90 MHz. Therefore, if the control is excited when $\Delta \nu = -90$ MHz the pulse intended to drive the transition $|0\rangle \rightarrow |e\rangle$ of the target qubit instead drives $|1\rangle \rightarrow |e\rangle$, with the reverse being true when $\Delta \nu = +90$ MHz. This is the reason behind the error spikes at $\pm90$ MHz. Their widths differ because of the differences in oscillator strengths between the intended transition and the transition being driven. 

TQ control-arbitrary gate operations with average errors of about $2\cdot 10^{-3}$ can be achieved over a large range of dipole-dipole shifts. However, since the blockade gate relies on the dipole blockade mechanism it only works well for relatively large shifts. In order to also utilize the qubits with weak coupling a so-called interaction gate is also examined, and is described below for a gate that adds a phase of $\pi$ to the $|00\rangle$ TQ state: 

\begin{enumerate}
    \item Excite both qubits simultaneously on their $|0\rangle \rightarrow |e\rangle$ transition, aiming to excite everything despite the dipole-dipole shift that occurs
    \item Wait a specific amount of time during which the $|ee\rangle$ TQ state evolves an additional phase due to the specific dipole-dipole shift between the two qubits
    \item Deexcite both qubits simultaneously on their $|0\rangle \rightarrow |e\rangle$ transition with an additional phase of $\pi$ on both transitions, thus deexciting the qubits on the same path as they were excited
\end{enumerate}

In order to excite both qubits despite the dipole-dipole shift that occurs, while still reducing the impact of ISD by minimizing the effect on non-qubit ions just outside the transmission windows, a hyperbolic-square-hyperbolic (HSH, also called sechscan) pulse shape \cite{Tian2011} is used instead of a cut Gaussian. The parameters for these pulses are optimized for each specific dipole-dipole shift, see Appendix \ref{app:optimization} for more information. The TQ gate error for this interaction gate when averaging over six initial states for each qubit, can be seen in Fig. \ref{fig:TQ_eval}b. The drastic increase in gate errors for $|\Delta\nu| > 7.5$ MHz arises since the optimization algorithm also takes heed for ISD, thus limiting the maximum frequency bandwidth of the pulses in order to not excite the non-qubit ions just outside the transmission windows. 

When utilizing both TQ gate types one can achieve an average TQ gate error ranging from $5\cdot 10^{-4} \rightarrow 3\cdot 10^{-3}$ as long as $|\Delta \nu|$ is larger than $0.1$ MHz and avoids the spikes in error when the shift coincides with the frequencies of other internal transitions, e.g., the spikes at $\pm90$ MHz seen in Fig. \ref{fig:TQ_eval}a.

In general the blockade gate is independent on the exact dipole-dipole shift, and the frequency bandwidth of the gate operation must be less than $|\Delta\nu|$ for the gate to be successful, something that can be both an advantage or a disadvantage. On the contrary, the interaction gate requires the knowledge of the exact shift and must have a frequency bandwidth that covers the dipole-dipole shift. For a given shift this means that the interaction gate can be made faster than the blockade gate, and therefore suffers less from decay and decoherence, at the cost of a larger frequency bandwidth. In this work the two gate types are used in different ranges of dipole-dipole shifts to utilize their strengths without suffering much from their disadvantages. In other words, the blockade gate is used for large dipole-dipole shifts where the frequency bandwidth of the pulses are limited by other constrains, e.g., reducing internal crosstalk or minimizing the risk of ISD occurring. Whereas the interaction gate is used for small dipole-dipole shifts where it is possible to utilize the faster gate operations without requiring a too high frequency bandwidth. 

% ---------------------------------------------------------------------------
\section{\label{sec:uncert}Uncertainties in system parameters}
In order to estimate how precisely one needs to know the system parameters of the qubits in order to achieve the low gate errors shown in Fig. \ref{fig:SQ_eval} and \ref{fig:TQ_eval}, we here investigate how sensitive the gate errors are toward changes in various parameters. 

\begin{figure*}
\includegraphics[width=\textwidth]{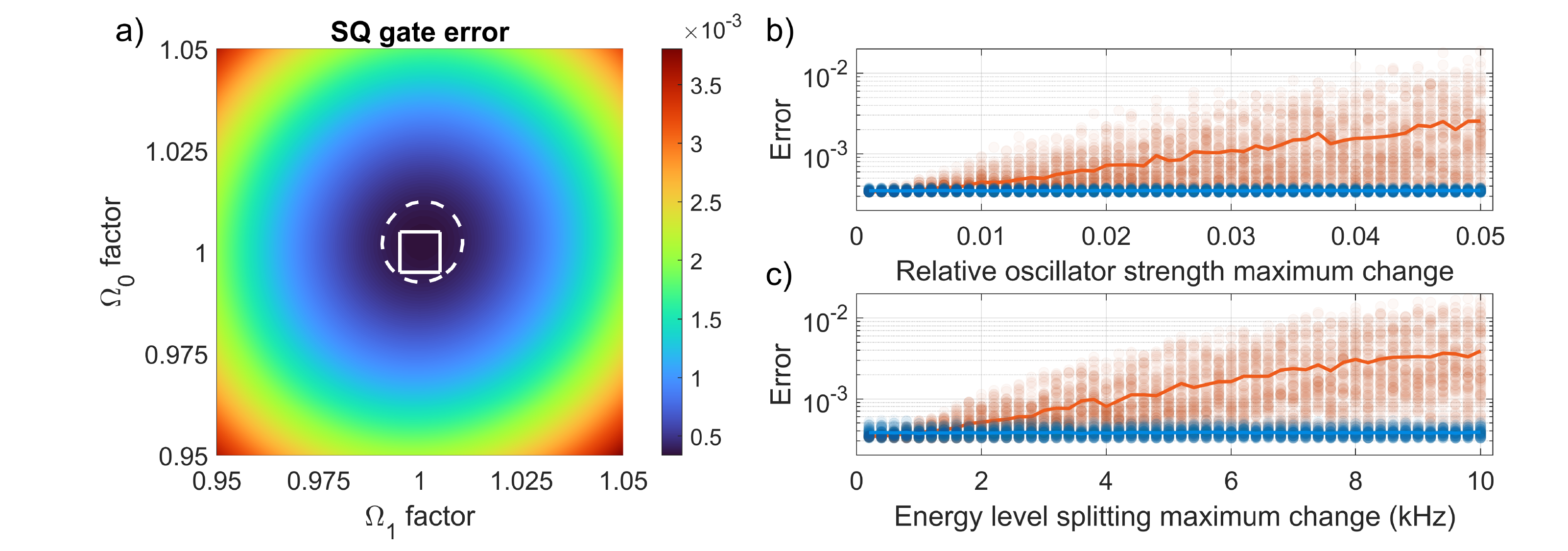}
\caption{\label{fig:SQ_uncertainties}Shows the SQ gate error as a function of a) multiplicative changes in the Rabi frequencies, $\Omega_0(t)$ and $\Omega_1(t)$, b) uncertainties in the relative oscillator strength values, and c) uncertainties in the energy level splittings. a) If the Rabi frequencies are allowed to fluctuate by up to $\pm0.5\%$ (corresponding to intensity fluctuations of roughly $\pm1\%$), which is indicated by the solid white box, the SQ gate error is still lower than $4\cdot 10^{-4}$, shown in the dashed white circle. b-c) The horizontal axes show how much (b) the relative oscillator strength values and (c) the energy level splittings, at most are allowed to differ from the parameters listed in Fig. \ref{fig:Enery_levels}. For each value on the horizontal axis, 100 randomizations are performed (half-transparent circles), and the average SQ gate error is calculated (solid lines). In the red data the gate pulses are constructed without knowledge of these random changes, and are therefore implemented with (b) the wrong Rabi frequencies and (c) the wrong resonant frequencies. This leads to larger SQ gate errors the more uncertain the system parameters are. However, in an experimental setting the Rabi and transition frequencies of the gate pulses can easily be fine-tuned to compensate for such uncertainties, at least for the two transitions that are being driven. This is simulated in the blue data, where (b) the Rabi frequencies only fluctuate within $\pm0.5\%$, and (c) the resonant frequencies are known to within at most $\pm1$ kHz, when compared to the ideal values of the randomized system. In these simulations, the uncertainties that change along the horizontal axes therefore mostly affect how the gate pulses off-resonantly interact with other optical transitions than the intended ones. Now the SQ gate error remains low regardless of the uncertainties in the system parameters.}
\end{figure*}

\begin{figure*}
\includegraphics[width=\textwidth]{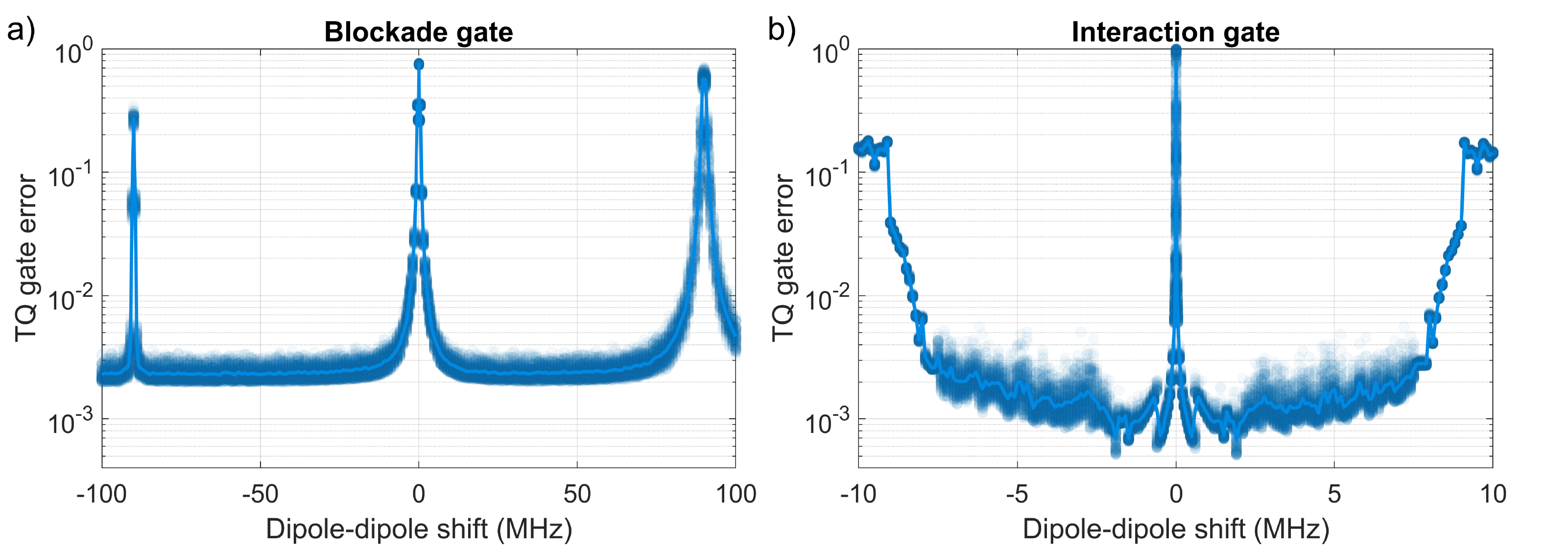}
\caption{\label{fig:TQ_uncertainties}Shows the TQ gate error as a function of the dipole-dipole shift when examining the impact of uncertainties in both the relative oscillator strengths (maximum uncertainty $0.05$) and level splittings (maximum uncertainty $10$ kHz) simultaneously. All gate pulses are optimized in the same way as the blue data in Fig. \ref{fig:SQ_uncertainties}b-c, i.e., when compared to the ideal values of the randomized system their Rabi frequencies only fluctuate by up to $\pm0.5\%$, and their resonant frequencies are known to within $\pm1$ kHz. 100 randomizations are performed for each shift (half-transparent circles), and the average TQ gate error is calculated (solid lines). This is simulated for a) a blockade gate performing a CNOT operation on the initial state $(|0\rangle + |1\rangle)\otimes(|0\rangle + i|1\rangle)$, and b) an interaction gate adding a $\pi$ phase to the $|00\rangle$ state performed on the initial state $(|0\rangle + |1\rangle)\otimes(|0\rangle + |1\rangle)$.}
\end{figure*}

Fig. \ref{fig:SQ_uncertainties} shows the SQ gate error as a function of fluctuations or uncertainties in a) Rabi frequencies, b) relative oscillator strengths, and c) level splittings. As can be seen one can still achieve high fidelity even when the system parameters are relatively uncertain as long as the frequencies of the two transitions that are being driven are known to at least $1$ kHz and that their Rabi frequencies are not fluctuating by more than roughly $\pm0.5\%$. Note that commercial laser amplitude stabilizers that can achieve fluctuations an order of magnitude less than this exist on the market today.

Note that these sensitivities might change if more sophisticated pulses are used, e.g., in DRAG schemes which compensate for internal crosstalk the knowledge of other transitions might be more important. Further note that while $1$ kHz resolution of the two transitions being driven is sufficient to keep the SQ gate error low, one might still want better resolution on the $|0\rangle \rightarrow |1\rangle$ frequency since the qubit dephase on a timescale roughly equal to 1/(frequency uncertainty) even when no gate operation is performed, unless dynamic decoupling pulses are used \cite{Viola1998, Fraval2005, Pascual-Winter2012}.

We can also use the data in Fig. \ref{fig:SQ_uncertainties}c to make some predictions about the effect of spectral diffusion. Reference \cite{Oswald2018} measured that narrow spectral holes in Eu:Y$_2$SiO$_5$ only increase their linewidth by roughly 1.5 kHz over 49 days, which when combined with the data of Fig. \ref{fig:SQ_uncertainties}c indicates that the error contribution due to long term spectral diffusion is low. Note, however, that the results obtained in reference \cite{Oswald2018} were for a bulk crystal when examining spectral holes that consist of large ensembles of ions. There might therefore exist other spectral diffusion processes which can only be seen on the single ion level or when examining other samples than bulk.

Another uncertainty investigation, this time for two specific TQ gate operations, can be found in Fig. \ref{fig:TQ_uncertainties}. The reason for not showing all combinations of initial states and gate operations was due to the unjustifiable time this would take to simulate. Here one can see that even though the TQ gate errors are slightly increased due to the uncertainties, they remain relatively low and qualitatively unchanged. 

If these precisions are too difficult to achieve, or if the added error due to the fluctuations are deemed too high, one can always change the cut Gaussian pulse shape into, e.g., an hyperbolic secant pulse shape. Thus trading faster gate operations for robustness against fluctuations in both Rabi frequencies and resonance frequencies.

% ---------------------------------------------------------------------------
\section{\label{sec:conc}Conclusion}
In conclusion, we have through simulations indicated that one can perform gate operations with low errors in rare-earth-ion-doped crystals; $2.1\cdot10^{-4}$ and $3.4\cdot10^{-4}$ for arbitrary single-qubit gates when either no ISD is present or when we take heed to minimize the effect of ISD, respectively. Furthermore, two-qubit gates with errors ranging from $5\cdot 10^{-4} \rightarrow 3\cdot 10^{-3}$ were possible over a broad range of dipole-dipole interaction strengths ($|\Delta \nu| > 0.1$ MHz). Such error rates lie beneath the threshold for error correction using surface codes \cite{Raussendorf2007, Fowler2009} and lies at the border of the threshold for CSS codes \cite{Steane2003}. Furthermore, the required knowledge of system parameters is investigated and shows that a reasonable precision of the parameters is sufficient to keep the errors low. Finally, throughout the article we discuss how gate and system parameters affect gate errors, the required frequency bandwidth per qubit, and the risk of ISD occurring, thus highlighting many of the trade-offs that one need to do when designing a quantum computer based on rare-earth-ion-doped crystals. 

% ---------------------------------------------------------------------------
\begin{acknowledgments}
We want to thank Klaus M{\o}lmer for the useful discussions. This research was supported by Swedish Research Council (no. 2016-05121, no. 2015-03989, no. 2016-04375, and 2019-04949), the Knut and Alice Wallenberg Foundation (KAW 2016.0081), the Wallenberg Center for Quantum Technology (WACQT) funded by The Knut and Alice Wallenberg Foundation (KAW 2017.0449), and the European Union FETFLAG program, Grant No.820391 (SQUARE).
\end{acknowledgments}

% ---------------------------------------------------------------------------
\appendix
\section{\label{app:simulations}Simulations}
All simulations were performed by evolving the Lindblad master equation \cite{Manzano2020} using MATLAB's explicit Runge-Kutta ode45 function \cite{Dormand1980, Shampine1997}. The Hamiltonian consisted of $6$ or $36$ energy levels when examining SQ or TQ gates, respectively. All system parameters, including the decay and decoherence terms, were added based on the parameters listed in Fig. \ref{fig:Enery_levels}. The relative and absolute tolerances for the ode45 function were kept at $10^{-6}$ for most simulations, which for a single gate operation results in global errors of roughly $10$ times that. The simulations in Fig. \ref{fig:Error_A_detuning_B} used lower tolerances of $10^{-11}$ to $10^{-9}$ depending on the detuning. 

\section{\label{app:optimization}Optimization of pulse parameters}
The optimizations of pulse parameters were performed using a combination of the GlobalSearch and fminsearch tools in MATLAB, where the initial guess was provided through a manual search of the parameter space. 

The SQ gate parameters $t_g$ and $\sigma$ from Eq. \ref{eq:Gaussian} were optimized for two different scenarios; without ISD or trying to minimize the risk of ISD. When optimizing the parameters without ISD the optimization tried to find the lowest average SQ error when performing the six gate operations on the six initial states listed in the main text. Here the optimization also included a parameter controlling the off-quadrature DRAG compensation, $\alpha_y$, as explained in the main text.

When taking heed for ISD the gate parameters were optimized by simultaneously trying to achieve two competing goals. First, find the lowest average SQ error as described above. Second, try to minimize the risk of ISD by doing as little as possible on ions detuned more than roughly $9$ MHz from the qubit, since this is the closest frequency from the qubit to an edge of the semipermanent transmission windows. If the effect of ISD is estimated by only looking at the impact on ions with one detuning, the pulses might be optimized to have a minimum in their excitation probability for only that detuning. Therefore, several detunings were evaluated between $-11$ MHz and $-9$ MHz. Note that one could equally well use only positive detunings, or a combination of positive and negative detunings, and that the range of detunings could be chosen differently. The important part is to evaluate multiple detunings over a wide enough frequency range so that the optimization doesn't produce gate parameters which have a local minimum in their effect on ions at the specified detunings. The two goals were weighted equally in the optimization procedure, i.e., the SQ gate error for the resonant case and the error of doing nothing on the detuned ions were calculated and the sum of these two errors were minimized in the optimization procedure. The resulting gate parameters indicated that DRAG compensation was not needed in this case where we take heed to minimize the effect of ISD, and was therefore not used in subsequent simulations using these pulse parameters.

The blockade control-arbitrary gates uses the same SQ gate parameters that take heed for ISD to perform the gate operation on the target qubit. However, new parameters were optimized for the excitation of the control qubit. The procedure was similar to that described above, except that the only gate operation was to excite/deexcite $|0\rangle \rightarrow |e\rangle$, and the initial states were the six superpositions both between $|0\rangle$ and $|1\rangle$ and between $|e\rangle$ and $|1\rangle$. 

The TQ interaction gate uses two hyperbolic-square-hyperbolic (HSH, also called sechscan) pulses \cite{Tian2011}, which are hyperbolic secant (sechyp) pulses \cite{Roos2004} with a linear frequency chirp added in the middle of the pulse, defined as follows, where $\Omega(t) = |\Omega(t)|e^{-i\phi(t)}$: 
\begin{equation}\label{eq:SechscanOmega}
    |\Omega(t)| = 
    \begin{cases}
      \Omega_0 \text{sech}(\beta (t - t_0)) & 0 \leq t < t_0\\
      \Omega_0 & t_0 \leq t \leq t_0 + t_\text{scan}\\
      \Omega_0 \text{sech}(\beta (t - t_0 - t_\text{scan})) & t_0 + t_\text{scan} < t \leq t_g \\
      0 & \text{otherwise}
    \end{cases}
\end{equation}

and
\begin{widetext}
\begin{equation}\label{eq:SechscanPhi}
    \phi(t) = 
    \begin{cases}
      -\mu \ln [\text{sech}(\beta (t-t_0))]-\frac{2\pi f_\text{scan}}{2}t & 0 \leq t < t_0\\
      \frac{2\pi f_\text{scan}}{2} (-t + \frac{(t - t_0)^2}{t_\text{scan}}) & t_0 \leq t \leq t_0 + t_\text{scan}\\
      -\mu \ln [\text{sech}(\beta (t - t_0 - t_\text{scan}))] + \frac{2\pi f_\text{scan}}{2} (-t_0 + (t - t_0 - t_\text{scan})) & t_0 + t_\text{scan} < t \leq t_g \\
      0 & \text{otherwise}
    \end{cases}   
\end{equation}
\end{widetext}

using
\begin{eqnarray}
    \beta = \frac{2\ln(1+\sqrt{2})}{t_\text{fwhm}} \nonumber\\
    \mu = \frac{\pi f_\text{width}}{\beta} \nonumber \\
    t_\text{scan} = \frac{2\pi f_\text{scan}}{\mu \beta^2} \nonumber \\
    t_0 = \frac{t_g - t_\text{scan}}{2} 
\end{eqnarray}

where $\Omega_0$ is the maximum Rabi frequency, $t_\text{fwhm}$ is the full-width-at-half-maximum of the intensity profile for the sechyp envelope, $f_\text{width}$ is the frequency width of the sechyp, $f_\text{scan}$ is the frequency range of the linear chirp which is applied over a duration of $t_\text{scan}$, and finally, $t_g$ is the cut-off duration of the pulse which lasts from $t = 0 \rightarrow t_g$. 

\begin{figure}
\includegraphics[width=\columnwidth]{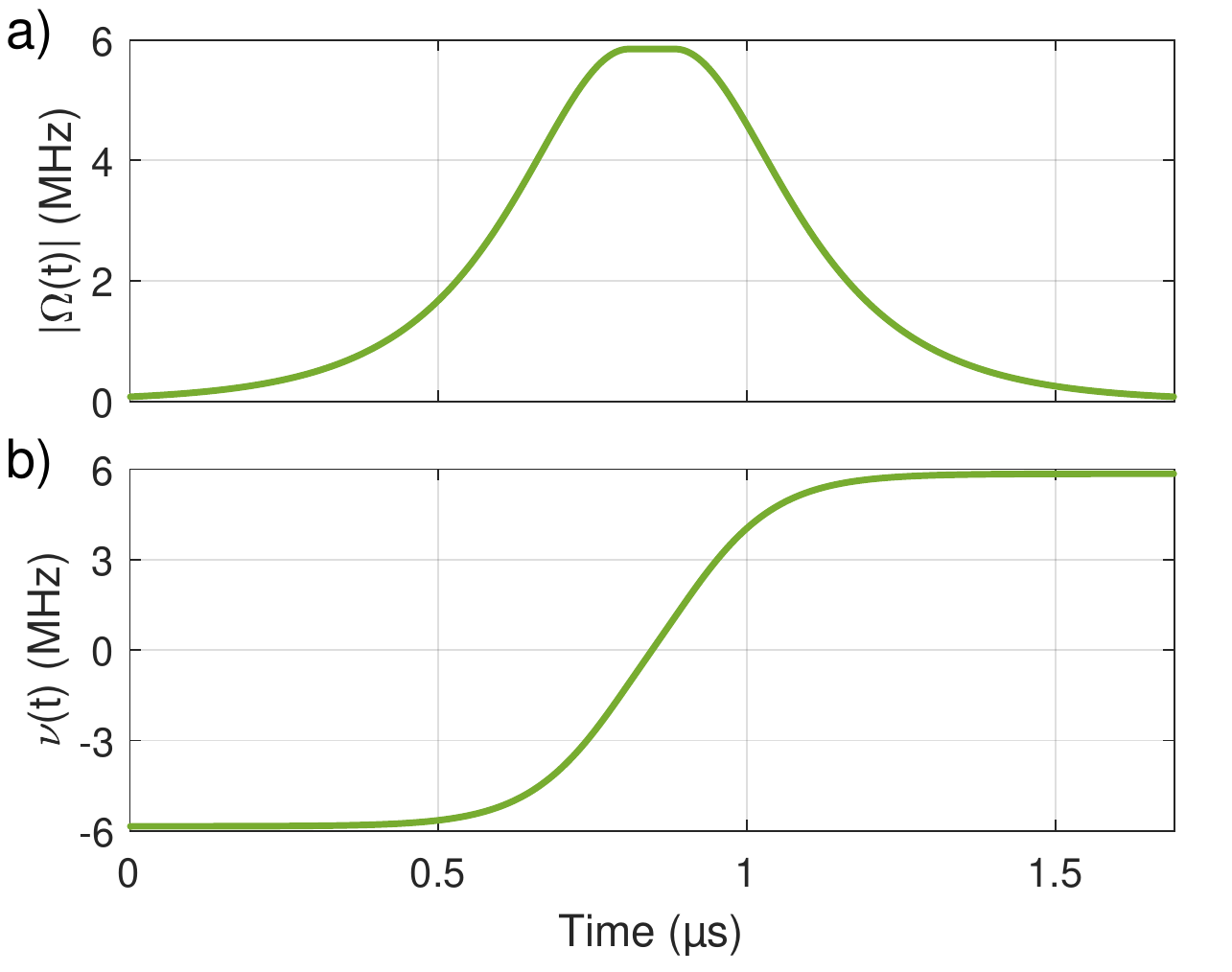}
\caption{\label{fig:SechscanPulseShape}a) The absolute Rabi frequency, where $|\Omega(t)|$ is given by Eq. \ref{eq:SechscanOmega}, and b) instantaneous frequency, where $\nu(t)=\frac{1}{2\pi}\frac{d\phi(t)}{dt}$ is calculated from $\phi(t)$ in Eq. \ref{eq:SechscanPhi}, of the sechscan pulse using parameters optimized for a dipole-dipole shift of $3$ MHz is shown. The optimized gate parameters for this specific shift are: $t_g \approx 1.7$ $\upmu$s, $t_{\text{fwhm}} \approx 0.28$ $\upmu$s, $f_{\text{width}} \approx 9.5$ MHz, $\Omega_0 \approx 5.8$ MHz, and $f_{\text{scan}} \approx 2.2$ MHz.}
\end{figure}

The pulse must excite both qubits regardless of the dipole-dipole shift that occurs, which means that each shift has a different requirement on the frequency bandwidth of the pulse. Therefore, new pulse parameters were optimized for each individual shift. Similarly as before, two competing goals were used in the optimization procedure. First, find the lowest error of the gate operation that adds a phase $\pi$ to the $|00\rangle$ TQ state performed on the initial state $(|0\rangle + |1\rangle) \otimes (|0\rangle + |1\rangle)$. Second, minimize the risk of ISD by doing as little as possible on ions detuned by $-11$ MHz $\rightarrow-9$ MHz from either qubit. Again, the two goals are weighted equally as described previously. The gate parameters vary, but have sechyp parameters of roughly $t_g = 1.5\pm0.5$ $\upmu$s, $t_{\text{fwhm}} = 0.32\pm0.14$ $\upmu$s, $f_{\text{width}} = 12\pm3.5$ MHz, and $\Omega_0 = 5.1\pm1.3$ MHz, with a linear chirp of $f_{\text{scan}} = 2.1\pm0.7$ MHz and $f_{\text{scan}} \approx 0$ MHz, for shifts lower and higher than $3.6$ MHz, respectively. The shape of the pulse when using the optimized parameters for the case of a dipole-dipole shift of $3$ MHz can be seen in Fig. \ref{fig:SechscanPulseShape}.

The phase of the final $|00\rangle$ TQ state evolves both when the qubits are being driven by the gate pulses, and during the wait when the qubit is in $|ee\rangle$. In order to determine how long wait duration is required for each shift, a simulation is first performed running the gate sequence without waiting between the two optical pulses. This allows us to determine the phase obtained due to only the qubits being driven. By comparing this phase to the desired phase of the TQ state, one can calculate the additional phase that should be added during the wait, which for a specific shift also gives you the wait duration. Note that the same wait duration must be used for all initial states of the two qubits. Therefore, the simulation performed in order to calculate the wait duration is only performed once, in our case on the initial TQ state $|0\rangle \otimes (|0\rangle + |1\rangle)$. 

% ---------------------------------------------------------------------------
\bibliography{Ref_lib}% Produces the bibliography via BibTeX.

\end{document}